\documentclass[12pt]{article}
\pdfoutput=1
\usepackage[sort&compress,square,comma,numbers]{natbib}
\usepackage{amsmath,amssymb,graphicx,slashed}
\usepackage{placeins}
\usepackage[tableposition=below]{caption}
\usepackage{longtable}
\usepackage{color}			
\usepackage{array}
\usepackage{verbatim}

\usepackage[
	colorlinks=true,
	citecolor=black,
	linkcolor=black,
	urlcolor=blue,
	hypertexnames=false]{hyperref}

\numberwithin{equation}{section} 

\def\tilde#1{\widetilde{#1}}

\newcommand{\arXiv}[2]{\href{http://arxiv.org/pdf/#1}{{\tt #2/#1}}}
\newcommand{\arXivold}[1]{\href{http://arxiv.org/pdf/#1}{{\tt #1}}}

\newcommand{\beq}{\begin{equation}}
\newcommand{\eeq}{\end{equation}}
\definecolor{MyRed}{rgb}{0.9,0.12,0.1}
\definecolor{MyBlue}{rgb}{0.1,0.12,0.9}

\begin{document}
\begin{titlepage}

\begin{center}

	{	
		\LARGE \bf 
		Little Conformal Symmetry
	}
	
\end{center}
	\vskip .3cm
	
	\renewcommand*{\thefootnote}{\fnsymbol{footnote}}


\begin{center}
		
		\bf
		Rachel Houtz\footnote{\tt \scriptsize
		 \href{mailto:houtz@ms.physics.ucdavis.edu}{houtz@ms.physics.ucdavis.edu},
		 $^\dag$\href{mailto:colwell@ms.physics.ucdavis.edu}{colwell@ms.phyics.ucdavis.edu},
		 $^\ddag$\href{mailto: terning@physics.ucdavis.edu}{terning@physics.ucdavis.edu},
		 },
		Kitran Colwell$^{\dag}$,
		%
		and John Terning$^{\ddag}$
\end{center}
	
	\renewcommand{\thefootnote}{\arabic{footnote}}
	\setcounter{footnote}{0}


\begin{center} 

	{\it Department of Physics, University of California, One Shields Ave., Davis, CA 95616}

\end{center}


\centerline{\large\bf Abstract}
\begin{quote}
We explore a new class of natural models which ensure the one-loop divergences in the Higgs mass are cancelled.
The top-partners that cancel the top loop are new gauge bosons, and the symmetry relation that ensures the cancellation arises at an infrared fixed point. Such a cancellation mechanism can, {\em a la} Little Higgs models, push the scale of new physics that completely solves the hierarchy problem up to 5-10 TeV. When embedded in a supersymmetric model, the stop and gaugino masses provide the cutoffs for the loops, and the mechanism ensures a cancellation between the stop and gaugino mass dependence of the Higgs mass parameter.
 \end{quote}

\end{titlepage}

\section{Introduction}

The discovery of a $125$ GeV Higgs boson by the CMS and ATLAS
experiments  fills in the final particle in the Standard
Model (SM). In order to achieve the measured Higgs mass, however, the
SM requires a fine-tuning of the mass parameter to cancel against
divergent loop contributions. Meanwhile, LEP constraints suggest that
higher dimension operators are suppressed by a scale around 10 TeV
\cite{LEP}. This ``little" hierarchy problem provides a
compelling motivation to study physics beyond the SM which can cancel
the divergent loop contributions and make the theory technically
natural. SUSY is the prime example of a natural theory, with stop
squark loops canceling the divergent part of the top loops, and
gauginos doing the same for the gauge loops. However, with the absence
(so far) of any kind of colored top-partners at the LHC there has been a great rush to abandon naturalness as a guiding principle in searching for extensions of the the SM. This has lead to an advance (retreat?) into anthropic landscapes, which seem to be highly non-predictive for particle physics.

Before we abandon naturalness, it would be good to have some idea of how many classes of natural theories exist that
have not yet been excluded by LHC data.  For example, Little Higgs theories provide an alternative class of natural theories where a global symmetry provides a top-partner which is a fermion  (rather than a scalar as happens in SUSY). There has also been recent progress finding new types of natural models in the class of ``neutral naturalness" models \cite{neutralnaturalness}, where a discrete symmetry provides a fermionic top-partner that is color neutral.

Here we examine yet another type of natural theory where the
top-partner is a gauge boson\footnote{A model with a gauge boson top
  partner has already appeared \cite{Cai:2008ss}, but in that case the
  gauge boson was a superpartner of the top quark.}. In order to
cancel the quadratic divergence in the top loop, the top partner gauge
boson needs to have a coupling that is related to the top Yukawa
coupling. In our scenario this relation between the couplings  is
accidental in that it is not the result of a symmetry of the
Lagrangian, but arises only at an infrared fixed point. As in
\cite{Tavares:2013dga}, we expect that this extra bit of conformal
symmetry can only cancel divergent diagrams at one-loop, since
conformal symmetry does not by itself ensure that scalar masses vanish. With a one-loop cancellation we can proceed as in Little Higgs models, and push the scale of new physics that completely solves the hierarchy problem up to 5-10 TeV.  For example, superpartners could have 10 TeV masses, so SUSY would ensure that the Higgs mass is kept below 10 TeV, and our new mechanism, which we will refer to as Little Conformal Symmetry, can ensure the cancellations that keep the Higgs below 1 TeV.

Of course, there have been many attempts to use conformal symmetry to
address the hierarchy problem \cite{conformalstuff}, but these
attempts typically stumble on imperfect cancellations
\cite{Tavares:2013dga,Csaki:1999uy}, or ultimately on the existence of
the Planck scale. The underlying problem, as we have said, is that
while the vanishing of scalar masses is necessary for the existence of
conformal symmetry, conformal symmetry does not, by itself, enforce a
vanishing scalar mass. However if we only need the cancellation to
work at one-loop, then there is still hope that conformal symmetry can
be useful. There is also a further problem that conformal symmetry cannot help with: what determines the cutoffs in the divergent loops?  This can only be answered in a theory where the cutoffs are calculable, so we will examine the case where Little Conformal Symmetry embeded in a supersymmetric model with gauge mediation where the ratio of stop and gaugino masses is fixed by gauge couplings.

This paper is organized as follows: in the next section we describe a simple toy model the illustrates the mechanism, then we turn to a more realistic model which does not assume that the different loop cutoffs are equal. Finally we present our conclusions and make suggestions for searching for a completely realistic model.

\section{A Toy Model}

\begin{figure}[h!]
\centering
$-im_H^2(p^2)=$\raisebox{-.16in}{\includegraphics[scale=.5]{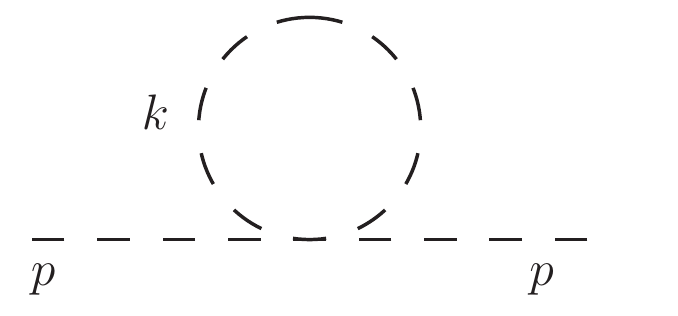}}$+\quad$\raisebox{-.43in}{\includegraphics[scale=.5]{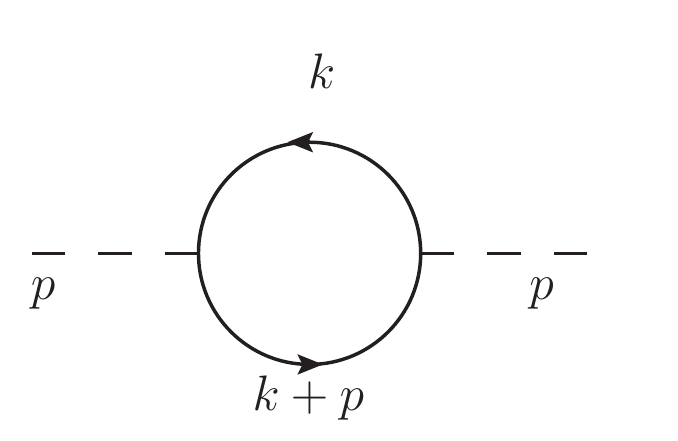}}$+\quad$\raisebox{-.16in}{\includegraphics[scale=.5]{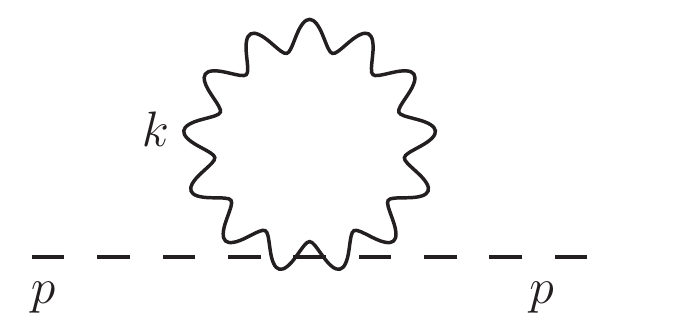}}\caption{Quadratically divergent diagrams that contribute to the Higgs mass}
\label{fig:divergence}
\end{figure}

The contributions to the quadratic divergence of the Higgs mass come
from the following diagrams shown in Fig.~\ref{fig:divergence}. 
Including only the leading fermionic contribution from the top quark
with $N_c$ colors, the quadratic correction to the Higgs mass is
\cite{Chaichian:1995ef}: 
\begin{align}
-im_H^2(0)=\left[6\lambda-2N_c \,y_t^2+3g_i^2C_2^i(H)\right]\int^{\Lambda}\frac{d^4k}{(2\pi)^4}\frac{1}{k^2}\label{QuadraticDivergence}
\end{align}
where $\lambda$ is the Higgs quartic coupling, $y_t$ is the top
Yukawa, the $i$ index runs over all the gauge couplings of
the Higgs, and $C_2^i(H)$ is the quadratic Casimir of the Higgs field
representation in the $i^{\text{th}}$ gauge group \cite{Slansky:1981yr}.
	
Long ago, Veltman \cite{Veltman:1980mj} suggested that there could be
a cancellation of these disparate 
contributions. However, no symmetry was found that would ensure the
cancelation, and the required top mass was the (then) almost
unimaginably large value of 69 GeV. A further problem with Veltman's
proposal is that if some 
new physics cuts off the integrals, then there is no guarantee that
the cutoffs of the three different types of loops would be the same.
In the context of SUSY, the top loop is cut off by the stop squark
mass, while the gauge loops are cut off by the gaugino masses, which
are typically not equal in models of dynamical SUSY breaking. 
	
Let us first consider how the top loop could be cancelled by a new non-Abelian gauge boson loop. For this to occur the Higgs would have to be embedded in a multiplet of a new gauge group.  For the Yukawa coupling to be gauge invariant, one or both of the left-handed and right-handed tops would also have to transform under the new gauge symmetry. Let us also assume for now that the cutoffs of the  two loop integrations are the same. (We will return to this point in the next section, where the cutoffs will be superpartner masses that will not be equal.) Ensuring the cancellation gives us a requirement that the top Yukawa coupling, $y_t$, is related to the new gauge coupling $g_N$. Surely no symmetry of a Lagrangian could force such a relation, but if the top Yukawa coupling is at an infrared fixed point that was determined by the value of the new gauge coupling, which itself is at an infrared fixed point, then there is a relation between the two couplings.

In order to construct a simple toy model that illustrates this mechanism, we will for now ignore the effects of the SM gauge groups and the quartic Higgs coupling. For concreteness let us embed the Higgs and the right-handed top in a fundamental and anti-fundamental of a $SU(N)$ gauge group.
The $\beta$-function for the top Yukawa coupling in this theory is
\cite{Machacek:1983tz,Arvanitaki:2004eu}: 
\beq
\label{topRGE}
\frac{d y_t}{d \ln \mu}=\frac{1}{16\pi^2}\left[\frac{1}{2}\left(2+n_D\right)+N_c\right]y_t^3-\frac{3}{16\pi^2}\left[C_2(t_R)+C_2(t_L)\right]g_N^2y_t
\eeq
where $C_2(F)$ is the quadratic Casimir of the representation of fermion $F$ 
under $SU(N)$, and  $n_D$ is the number of Higgs doublets, so in the
SM $N_c=3$ and $n_D=1$ and the first coefficient reduces to the standard result
$9/32\pi^2$. In our toy model
$C_2(t_L)=0$ and $n_D=N$, so a Yukawa fixed point occurs when \cite{Hill:1980sq} 
\beq
0=\left(4+\frac{N}{2}\right)y_{t*}^2-3\, C_2(t_R)g_{N*}^2
\label{toyyukawafixedpoint}
\eeq
Cancelling the
quadratic divergence simultaneously requires
\beq
0=-2N_c \, y_{t*}^2+3 \,C_2(H) g_{N*}^2~,
\label{CancelQuadDiv}
\eeq
which gives us a relation between the Casimirs of the Higgs field and
right-handed top quark:
\beq
\frac{C_2(H)}{C_2(t_R)}=\frac{12}{8+N}~,
\eeq
and since the $t_R$ and the Higgs are in conjugate $SU(N)$
representations the solution is $N=4$. 

For the toy model to be consistent we also need the $SU(4)$ gauge coupling to be at a fixed point.  The two-loop gauge $\beta$-function is
\beq
\frac{d g_N}{d \ln \mu}=-\frac{1}{16 \pi^2}\left(b \, g_N^3+c \,
g_N^5+d \, g_N y_t^2\right)~.
\label{gaugebetafn}
\eeq
When the gauge group is asymptotically free, i.e. $b>0$, then often $c<0$, and if $b$ is small there is a perturbative Banks-Zaks fixed point \cite{BanksZaks} for $y_t=0$. There is no general theorem determining whether there are fixed points for $y_t\ne0$, but they can be easily found by scanning over the possible gauge representations of the matter fields.
Generically there are multiple solutions for fixed points of the coupling $g_N$ at the fixed point $y_{t*}$, depending on the matter content of the gauge theory, we will call them $g_{N*i}$. In order to make an interesting model we would arrange $SU(4)$ to break at some scale around $\Lambda \sim$ 5-10 TeV; that is, parts of the extended $t_R$ and Higgs multiplets can get masses at this scale, while the components corresponding to the SM $t_R$ and Higgs remain light. The gauge bosons could have masses somewhere between 1 TeV and $\Lambda$. Given the measured value of $y_t$ we could run it up towards the UV, and at each RG scale $\mu_i$ where $y_t(\mu_i)$ satisfies Eq.~(\ref{toyyukawafixedpoint}) with $g_{N*}=g_{N*i}$ we have a possible consistent model.


\section{A More Realistic Toy Model}

If we try to directly apply the mechanism of the previous section to the SM, we immediately run into a problem: the QCD contribution to top Yukawa $\beta$ function (\ref{topRGE}) is much larger that the Yukawa contribution. This implies that the new gauge group would not dominate at a fixed point and this would spoil our cancellation.  In order to get around this we can
embed color $SU(3)_c$ in $SU(N)$, and take both $t_L$ and $t_R$ to
transform under conjugate representations of $SU(N)$. Going further we can take a semi-simple gauge group $SU(N)\times SU(N')$ with the
understanding that $SU(3)_c$ is the diagonal subgroup of
$SU(3)_L\times SU(3)_R\subset SU(N)\times SU(N')$. We will also assume
that the top quarks are charged under an additional global $SU(M)$
symmetry to enhance the otherwise small top loop contribution to both
(\ref{topRGE}) and (\ref{CancelQuadDiv}).

In order to connect to the pheonomenology of the SM and ensure anomaly
cancellation for the $SU(N)$ and $SU(N')$ gauge fields, we must
introduce the spectator fermions $b_R$ that contain the right-handed bottom quark. For now we will only consider
Yukawa couplings that give masses to the top quark. A summary of the charge assignments for the third generation
quarks and Higgs fields consistent with anomaly cancellation are given
in Table \ref{charges} (a second choice that yields asymptotic freedom
and anomaly cancellation interchanges the $b_N$ and $b_M$ representations).

\begin{table}[h]
\centering
\begin{tabular}{>{$}l<{$}|*{2}{>{$}c<{$}}||*{1}{>{$}c<{$}}}
 	  & SU(N) 	    & SU(N')	      & SU(M) \\
\hline
\rule{0pt}{3ex}
\bar{t}_R & \overline{\Box} & 1 	      & \overline{\Box}	\\
H_u	  & \Box	    & \overline{\Box} & 1		\\
Q_L	  & 1 		    & \Box	      & \Box		\\
\bar{b}_N & \overline{\Box} & 1		      & 1		\\
\bar{b}_M & 1		    & \overline{\Box} & 1		\\
\end{tabular}
\caption{Top and Higgs charges under a bifundamental Higgs}
\label{charges}
\end{table}

\begin{figure}[h!]
\centering
$\displaystyle\frac{dy_t}{d\ln\mu}=$
\raisebox{-.36in}{\includegraphics[scale=.6]{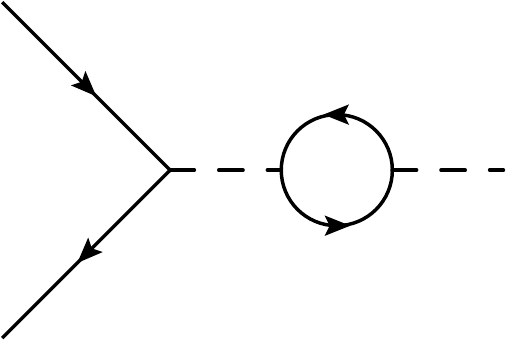}}
$\quad+\quad$\raisebox{-.36in}{\includegraphics[scale=.6]{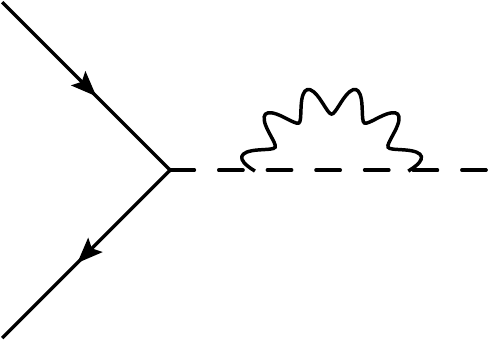}}
$\quad+\quad$\raisebox{-.36in}{\includegraphics[scale=.6]{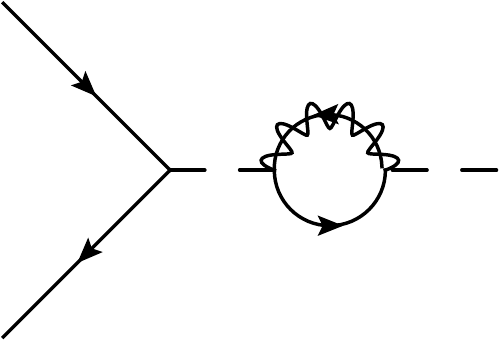}}\\
$\quad+\quad$\raisebox{-.36in}{\includegraphics[scale=.6]{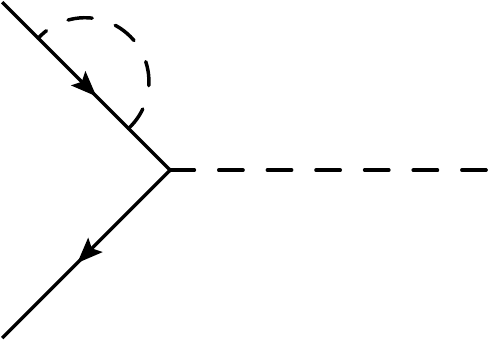}}
$\quad+\quad$\raisebox{-.36in}{\includegraphics[scale=.6]{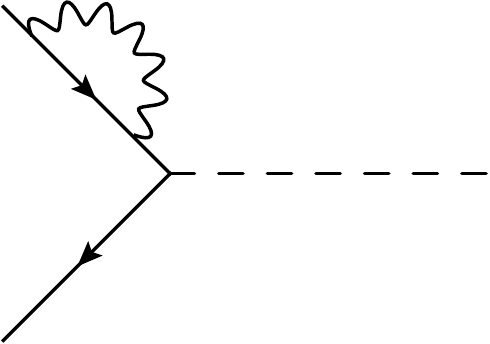}}
$\quad+\quad$\raisebox{-.36in}{\includegraphics[scale=.6]{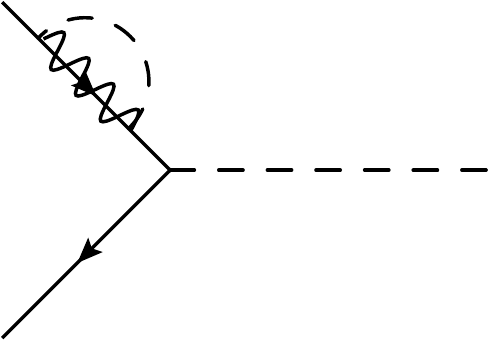}}\\
$\quad+\quad$\raisebox{-.36in}{\includegraphics[scale=.6]{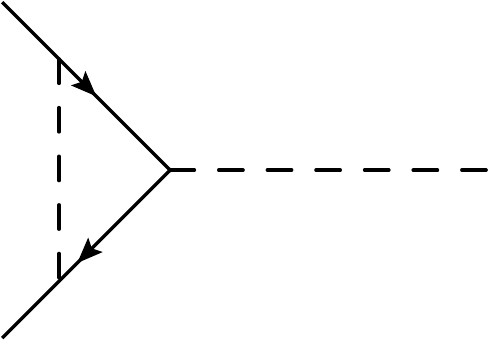}}
$\quad+\quad\text{$\bar{t}_R$ leg corrections}$
\caption{$\beta$ function for top Yukawa coupling}
\label{fig:topyuk}
\end{figure}

The one-loop running of a Yukawa coupling can be computed from the
diagrams in Figure \ref{fig:topyuk}. 
The index structure of our theory causes the 1PI contribution to
vanish; defining $\alpha_t\equiv y_t^2/4\pi$ and $\alpha_{N,N'}\equiv
g_{N,N'}^2/4\pi$, and using
the results of Machacek and Vaughn \cite{Machacek:1983tz} we can easily find
the one-loop beta function for the high-energy theory:
\beq
\frac{d\alpha_t}{d\ln\mu}=\frac{\alpha_t}{2\pi}\Big[\left(2N'+N+M\right)\alpha_t-4C_F\alpha_N-4C_F'\alpha_{N'}\Big]~.
\label{yukbeta}
\eeq

The cancellation condition (\ref{CancelQuadDiv}) with the more general 
gauge representations of the top and Higgs fields becomes:
\beq
0=2M\Lambda_t^2\alpha_t-3C_F\Lambda_N^2\alpha_N-3C_F'\Lambda_{N'}^2\alpha_{N'}~.
\label{QuadDivgen}
\eeq
Here we also account for the fact that the cutoffs for the integrals in the
quadratic divergence are in general different, and denote them as
$\Lambda_t$, $\Lambda_N$, and $\Lambda_{N'}$. In the context of SUSY,
$\Lambda_t$ is proportional to the stop mass, while the gauge cutoffs
are set by gaungino masses. To find a relation
between the cutoffs, we would need the details of the SUSY-breaking
mechanism.

A simple model for producing squark and gaugino masses is gauge
mediation \cite{Gaugemediation}. These models assume that there is Goldstino multiplet $X$ with a Yukawa coupling to messenger charged under the SM gauge groups and aVEV 
\beq
\langle X\rangle=M_{UV}+\theta^2\mathcal{F}~.
\eeq
This yields gaugino masses at one-loop given by
\beq
M_N=\frac{\alpha_N}{4\pi}N_m\frac{\mathcal{F}}{M_{UV}},\quad
M_{N'}=\frac{\alpha_{N'}}{4\pi}N_m'\frac{\mathcal{F}}{M_{UV}}~,
\label{gauginomass}
\eeq
with $N_m=2\sum T(R)$ the sum of indexes of the messengers. The stop masses are
\beq
\tilde{m}_R^2=2C_2(t_R)\frac{\alpha_N^2}{16\pi^2}N_m\left(\frac{\mathcal{F}}{M_{UV}}\right)^2,\quad
\tilde{m}_L^2=2C_2(t_L)\frac{\alpha_{N'}^2}{16\pi^2}N_m'\left(\frac{\mathcal{F}}{M_{UV}}\right)^2~,
\label{stopmass}
\eeq
so we expect the cutoffs to be
\beq
\Lambda_t^2=\frac{1}{2}\left(\tilde{m}_R^2+\tilde{m}_L^2\right)\ln\left(\frac{\Lambda^2+\tilde{m}_t^2}{\tilde{m}_t^2}\right),\quad
\Lambda_{N}^2=M_{N}^2\ln\left(\frac{\Lambda^2+M_{N}^2}{M_{N}^2}\right)~,
\label{cutoffs}
\eeq
for a UV scale $\Lambda$, and an equivalent expression for $\Lambda_{N'}$. The logarithmic factors are equal to leading
order in the gauge coupling, and may be cancelled in (\ref{QuadDivgen}), since the difference corresponds to a higher 
loop effect. We will also assume that the gauge coupling is at its fixed point up to the scale $\Lambda$; this may not be the case in a fully realistic model, but again
deviations from this limit correspond to higher order loop effects \cite{Tavares:2013dga}. 

In order to prevent a Higgs soft-mass at the 10 TeV scale, like the stop squarks, a
combination of chiral and vector-like messengers \cite{Nardecchia:2009nh} may be
needed in the high energy theory; due to the opposite signs of their
contributions to the gaugino masses, a judicious choice of
representations could allow for a light Higgs mass. We are of course primarily concerned with contributions to the Higgs mass from physics below the 10 TeV scale.

The new gauge coupling $\beta$ functions at two
loops are:
\begin{align}
\frac{d\alpha_N}{d\ln\mu}&=-\frac{\alpha_N^2}{2\pi}\left(b_N+c_N\frac{\alpha_N}{4\pi}+d_N\frac{\alpha_t}{4\pi}+e_{NN'}\frac{\alpha_{N'}}{4\pi}\right)\label{gaugeRGE1}\\
\frac{d\alpha_{N'}}{d\ln\mu}&=-\frac{\alpha_{N'}^2}{2\pi}\left(b_{N'}+c_{N'}\frac{\alpha_{N'}}{4\pi}+d_{N'}\frac{\alpha_t}{4\pi}+e_{N'N}\frac{\alpha_N}{4\pi}\right)
\label{gaugeRGE2}
\end{align}
The coefficients in (\ref{gaugeRGE1}) and (\ref{gaugeRGE2}) are
sensitive to the matter content of the UV theory, and ensuring
consistency between the three fixed point conditions and the
cancellation of the quadratic divergence requires specific choices of
representations and multiplicities of UV field content. 

\begin{figure}[h!]
\centering
\includegraphics[scale=.6]{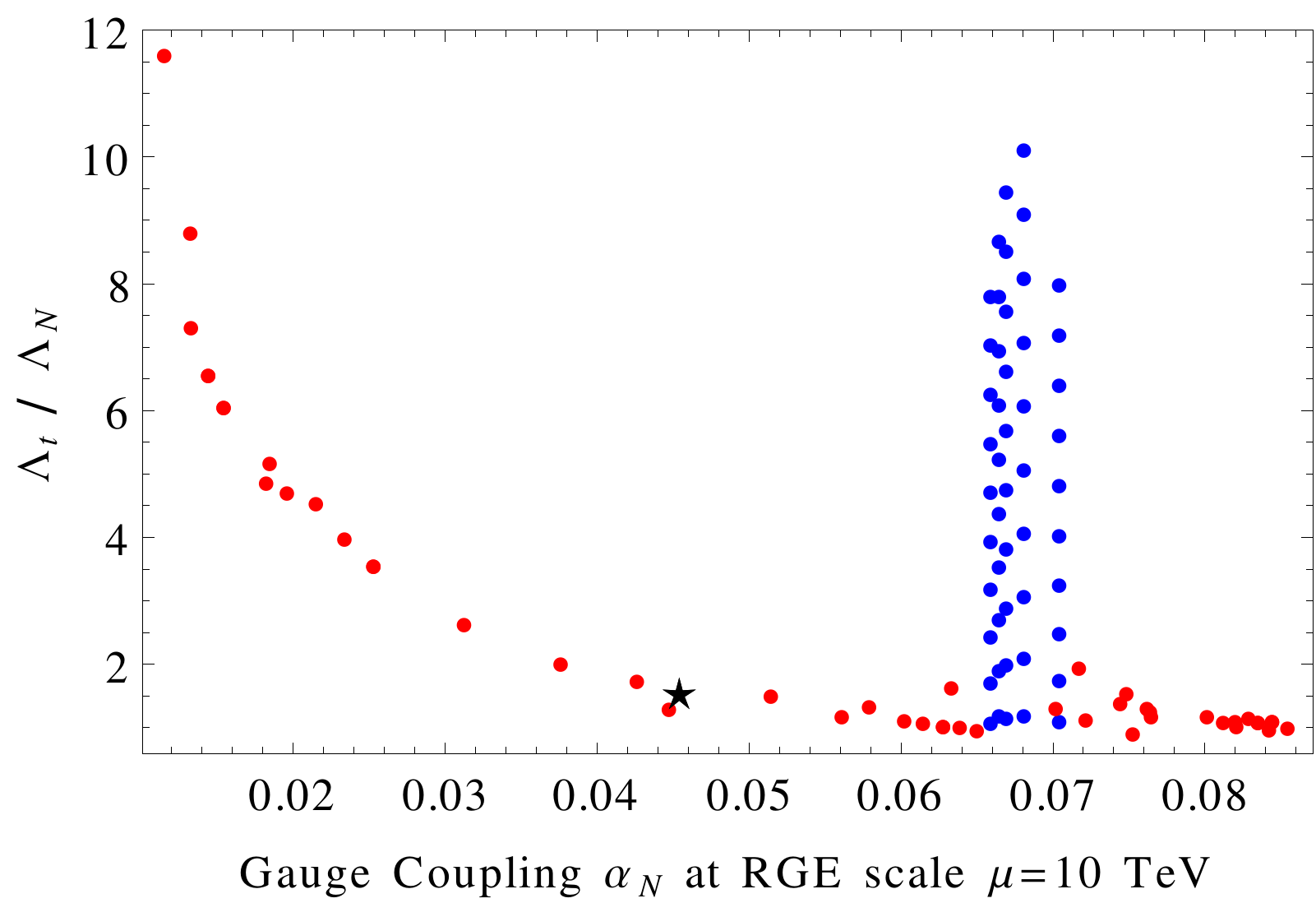}
\caption{This figure shows the cutoff ratios versus $SU(N)$ gauge coupling for various UV theories. The points in blue correspond to theories shown in Table \ref{UVmatter1} that satisfy Eq.~(\ref{alphabreaking}). The points in red correspond to theories shown in Table \ref{UVmatter3} that match the SM couplings, Eq.~(\ref{topexpt}). The black star represents an example theory whose cutoffs match a gauge-mediated SUSY-breaking scenario.}
\label{fig:scatter}
\end{figure}


At the scale $\Lambda_{IR}$ where
$SU(N)\times SU(N')\to SU(3)_c$, the gauge couplings must obey
\beq
\frac{1}{\alpha_3}\geq\frac{1}{\alpha_N}+\frac{1}{\alpha_{N'}}~.
\label{alphabreaking}
\eeq
There is tension between needing $\alpha_{N,N'}>\alpha_3$ to satisfy
(\ref{alphabreaking}) and having the $\alpha_t\sim\alpha_{N,N'}$ found
via (\ref{yukbeta}) or (\ref{QuadDivgen}) be small 
enough to run up to the SM value of the top Yukawa coupling.
Current measurements of the strong coupling constant, the top quark
mass, and Higgs VEV \cite{PDG} yield
\beq
\alpha_3^{\overline{MS}}(m_Z)=0.1185\pm 0.51\%,\quad
\alpha_t^{\overline{MS}}(m_t)=0.06721\pm 5.65\%~,
\label{topexpt}
\eeq
so we may choose parameters such that (\ref{alphabreaking}) or
(\ref{topexpt}) is satisfied, but not both. Table \ref{UVmatter1} of the
appendix summarizes theories with gauge couplings larger than
$\alpha_3$ at $\Lambda_{IR}=1$ TeV, while Table \ref{UVmatter2} summarizes
those that match the SM value of $\alpha_t$ at 1 TeV to within 5\%.
Figure \ref{fig:scatter} shows the ratio of the top-loop cutoff to the
$SU(N)$ gauge-loop cutoff for these theories. We made the
simplifying assumtion that
$\Lambda_N/\alpha_N=\Lambda_{N'}/\alpha_{N'}$ (or a rational multiple thereof) as in gauge mediation.
In the class of models studied we generically require a rather large
global $SU(M)$ symmetry for the top quarks, or a large running of
$\alpha_N$ or $\alpha_{N'}$ from the TeV scale to the 10 TeV scale.

While $t\bar{t}$ resonance experiments will be sensitive to additional
gauge symmetries, even greater experimental constraints come
from precision electroweak measurements \cite{Amaldi:1991cn}, which
would rule out the addition of many 
additional $SU(2)_L$ doublets like our extra top quark multiplets unless they have vector-like masses
that do not require a Higgs VEV. Any
solution to the Higgs mass problem should also include a mechanism for
sending $\alpha_2$ to a fixed point as well, a feature we have ignored. 
In this respect, a realistic model will probably look quite different from our toy model.

\section{Conclusions}

We have seen that new gauge interactions that
couple only to top quarks and the Higgs field 
allow for a cancellation of the top-loop quadratic divergence in the Higgs mass
at one-loop. To enforce this condition, the SM must be embedded in a UV theory (above a 5-10 TeV threshold) with fixed points for
the gauge couplings and the top Yukawa coupling. In a supersymmetric context, with calculable stop and gaugino masses, this can lead to
the cancellation of the stop and gaugino mass dependence in the Higgs mass.

Since the cancellation depends on the top and gauge loop cutoffs, in order to search for a fully realistic model one also has know the form of the cutoff. For supersymmetric theories this means that one needs to include a mechanism to mediate supersymmetry breaking 
to the supersymmetric SM sector. In this sense the mechanism is highly constrained, which also means that it is highly predictive.

We have not addressed how to account for  the divergences from the $SU(2)_L\times U(1)_Y$ gauge and Higgs quartic
couplings. It is possible that this could be addressed by additional fixed point conditions, or another mechanism altogether; for example
the theory could possess light (TeV scale) winos, binos, and higgsinos. In order to preserve the fixed point structure however, the $U(1)_Y$
would have to be embedded in a non-Abelian group.

In general, the cancellation of one-loop quadratic divergences in the Higgs mass does not require the existence of IR fixed points, we can also envisage that the ratios of couplings $\alpha_N/\alpha_t$ approaches a fixed value\footnote{We thank Hsin-Chia Cheng for pointing out this possibility.}. This is much less restrictive than enforcing IR fixed points for both couplings simultaneously, and provides hope that more realistic models can be constructed.

\appendix
\section*{Acknowledgments}
\setcounter{equation}{0}
\setcounter{footnote}{0}

We thank Roni Harnik for suggesting the title. We thank Hsin-Chia Cheng, Csaba Cs\'aki, Markus Luty, Stephen Martin, and Martin Schmaltz for helpful discussions. JT thanks the Aspen Center for Physics, where this work was initiated. This research was supported in part by the DOE under grant DE-SC-000999 and the NSF under grant PHY1066293.

\section{UV Matter Content}

UV matter content and gauge parameters for theories with
$\alpha_{N,N'}>\alpha_3$ at $\Lambda_{IR}=1$ TeV are given in Table 
\ref{UVmatter1}. Here a gauge group of $SU(N)\times SU(N')$ is assumed, with
a global symmetry group for top quarks of $SU(M)$. Weyl fermion
numbers $n_{ij}$ of UV matter content are scanned over with
$i,j=1,F,A$ corresponding to singlets, fundamentals, or adjoints under
$SU(N)$ and $SU(N')$. We assume SUSY in the UV theory to set the
number of real scalars as $s_{ij}=4n_{ij}$. Theories are listed in order of
ascending relative error in Eq.~(\ref{topexpt}).

Tables \ref{UVmatter2} and \ref{UVmatter3} list theories that match the running of $\alpha_t$ to (\ref{topexpt}). Table \ref{UVmatter2} lists the top five minimal matter content scenarios that satisfy Eq.~(\ref{topexpt}). Table \ref{UVmatter3} lists the theories that match most closely with the gauge-mediated SUSY breaking scenario described in Section 3.

All candidate
theories have $n_{A1},\,n_{1A},\,n_{FF},\,n_{AF},\,n_{FA},\,n_{AA}= 0$ due to their large impact
on the $\beta$-functions of the couplings.

\begin{table}[h!]
\centering
\begin{tabular}{*{3}{>{$}c<{$}}|*{3}{>{$}c<{$}}|*{5}{>{$}c<{$}}}
 N & N'& M & \alpha_N(\Lambda_{UV}) 
		    & \alpha_{N'}(\Lambda_{UV}) 
			    & \alpha_t(\Lambda_{UV}) 
				      & n_{b_N} 
					   & n_{b_M} 
				 	      &n_{F1} 
						  & n_{1F} 
				 		      \\
\hline 
 6 & 6 & 8 & .06602 & .06602 & .05925 & 4 & 4 & 4 & 0 \\
 4 & 4 & 4 & .06658 & .06658 & .06242 & 4 & 0 & 3 & 3 \\
 4 & 4 & 5 & .07053 & .07053 & .06225 & 3 & 2 & 3 & 1 \\
 6 & 7 & 9 & .06704 & .07631 & .06306 & 5 & 6 & 2 & 0 \\
 6 & 8 & 10& .06821 & .08040 & .06444 & 6 & 8 & 0 & 0 \\ 
\end{tabular}
\caption{UV Matter satisfying Eq.~(\ref{alphabreaking}) that have the smallest error in Eq.~(\ref{topexpt})}
\label{UVmatter1}
\end{table}

\FloatBarrier

\begin{table}[h!]
\centering
\begin{tabular}{*{3}{>{$}c<{$}}|*{3}{>{$}c<{$}}|*{5}{>{$}c<{$}}}
 N & N'& M & \alpha_N(\Lambda_{UV}) 
		    &\alpha_{N'}(\Lambda_{UV}) 
			     & \alpha_t(\Lambda_{UV}) 
				      &n_{b_N} 
					  & n_{b_M} 
					      & n_{F1} 
						  & n_{1F}\\
\hline
 5 & 4 & 5 & .04055 & .07286 & .05199 & 3 & 0 & 6 & 1 \\
 4 & 5 & 7 & .07947 & .04340 & .04822 & 3 & 6 & 1 & 0 \\
 4 & 5 & 6 & .07538 & .04164 & .04825 & 4 & 4 & 1 & 2 \\
 4 & 5 & 5 & .07128 & .03987 & .04828 & 5 & 2 & 1 & 4 \\
 5 & 4 & 4 & .03865 & .06844 & .05202 & 4 & 2 & 6 & 1 
\end{tabular}
\caption{Minimal UV Matter satisfying Eq.~(\ref{topexpt})}
\label{UVmatter2}
\end{table}

\FloatBarrier

\captionsetup[longtable]{skip=.5cm}
\begin{longtable}[h!]{c|c|c|c|c|c|c|c|c|c}
N &N'&M &$\alpha_N(\Lambda_{UV})$
		&$\alpha_{N'}(\Lambda_{UV}) $
			& $\alpha_t(\Lambda_{UV})$ 
				&$n_{b_N} $
					& $n_{b_M}$ 
						&$ n_{F1} $
							& $n_{1F}$\\
\hline
8 &8 &4 &.04547	&.04547	&.05115	&12	&8	&5	&5	\\
9 &9 &4 &.01336	&.07497	&.05065	&14	&10	&7	&4	\\
10&10&9 &.01334	&.08300	&.04893	&11	&2	&8	&4	\\
9 &9 &4 &.07497	&.01336	&.05065	&14	&10	&4	&7	\\
8 &9 &10&.08556	&.02270	&.04864	&8	&4	&2	&5	\\
10&8 &9 &.04486 &.05387 &.04962 &7	&2	&10	&1	\\
10&7 &9 &.08133	&.01133	&.05351	&5	&2	&10	&0	\\
10&8 &10&.08361	&.008507&.04971	&6	&0	&8	&3	\\
6 &5 &7 &.06512 &.04314 &.05103 &2	&2	&6	&0	\\
10&10&6 &.07631 &.02183 &.05398 &14	&8	&4	&7	\\
9 &8 &7 &.01970 &.07990 &.05027 &9	&4	&9	&2	\\
9 &7 &6 &.05623 &.03623 &.05160 &8	&6	&9	&1	\\
8 &10&9 &.08219 &.02859 &.05029 &11	&2	&0	&9	\\
10&10&10&.01453 &.08454	&.04904 &10	&0	&8	&4	\\
10&9 &10&.06290	&.03867	&.05086	&8	&0	&7	&4	\\
10&9 &9 &.08212	&.01825	&.05271	&9	&2	&6	&5	\\
8 &9 &6 &.02161	&.07734 &.05360 &12	&4	&4	&5	\\
8 &10&3	&.06343	&.03595	&.05519	&17	&10	&0	&10	\\
8 &6 &2 &.07184	&.005005&.05409	&10	&12	&8	&1	\\
7 &8 &6 &.07228 &.02734	&.04903	&10	&2	&2	&8	\\
8 &7 &10&.08435	&.01834	&.04938	&4	&4	&6	&0	\\
8 &8 &6	&.07658	&.02487	&.05326	&10	&4	&4	&6	\\
8 &9 &7 &.02350	&.07889	&.05372	&11	&2	&4	&6	\\
10&9 &8 &.06033	&.03535	&.05064	&10	&4	&7	&4	\\
9 &8 &4 &.05156	&.04232	&.05459	&12	&10	&7	&3	\\
6 &5 &5 &.06157	&.03688	&.05107 &5	&2	&6	&2	\\
7 &6 &4 &.07030	&.01650	&.05029	&8	&6	&6	&3	\\
6 &7 &7 &.06399 &.04243 &.04920 &7	&2	&2	&5	\\
10&10&10&.08454	&.01453	&.04904	&10	&0	&4	&8	\\
4 &5 &6 &.07538	&.04164	&.04825	&4	&4	&1	&2	\\
10&9 &4 &.03772 &.04939	&.05078	&14	&12	&8	&3	\\
8 &10&3 &.01164	&.07293	&.05249	&17	&10	&2	&8	\\
8 &6 &4 &.07456	&.01109	&.05432	&8	&8	&8	&1	\\
8 &8 &10&.03137	&.08442	&.05364	&6	&4	&6	&0	\\
9 &7 &7 &.08026	&.009409&.05186	&7	&4	&8	&2	\\
8 &10&6 &.04275 &.05681	&.05289	&14	&4	&1	&9	\\
6 &8 &6 &.01555 &.07655 &.04954	&10	&0	&1	&8	\\
6 &5 &3 &.05802 &.03062 &.05110	&7	&6	&6	&2	\\
8 &9 &8 &.02540 &.08045	&.05383	&10	&0	&4	&6	\\
6 &8 &7 &.01835	&.07780	&.04963	&9	&2	&1	&6	\\
9 &8 &6 &.01857	&.07775	&.05015	&10	&6	&9	&2	\\
9 &10&6 &.07652	&.02391	&.05239	&14	&6	&2	&9	\\
7 &9 &5 &.06643	&.03463	&.05089	&13	&4	&0	&10	\\

\caption{
UV matter content satisfying Eq.~(\ref{topexpt}), listed in order of how closely these theories match a gauge-mediated SUSY breaking scenario}
\label{UVmatter3}
\end{longtable}

\FloatBarrier

\end{document}